\documentclass[12pt]{article}

\usepackage{amsfonts}
\usepackage{amssymb}
\usepackage{amscd}

\def\D{\hbox{D\kern-.73em\raise.25ex\hbox{-}\raise-.25ex\hbox{ }}}
 \def\d{\hbox{d\kern-.33em\raise.75ex\hbox{-}\raise-.75ex\hbox{}}}

\def\GGG{\frak G }
\def\gr3{\GGG\,(\SSS_3)}

\def\gr2{\GGG\,(\SSS_2)}

\def\SSS{\frak S}

\def\al{{\alpha}}
\def\bet{{\beta}}

\def\gam{{\gamma}}

\def\vp{\vspace}
\def\hp{\hspace}
\def\ed{\end{document}}
\def\beq{\begin{equation}}
\def\eeq{\end{equation}}
\def\bea{\begin{eqnarray}}
\def\eea{\end{eqnarray}}
\def\ba{\begin{array}}
\def\ea{\end{array}}
\def\bi{\begin{itemize}}
\def\ei{\end{itemize}}

\def\noi{\noindent}
\def\nn{\nonumber}

 \def\dst{\displaystyle}
\newcommand{\bp}{\noindent\begin{minipage}[c]}
\newcommand{\ep}{\end{minipage}}

\begin{document}
 \baselineskip=11pt

\title{\Large Noncommutative Classical and Quantum Mechanics for
Quadratic Lagrangians (Hamiltonians)\hspace{.25mm} }
\author{\bf{Branko Dragovich}\hspace{.25mm}\thanks{\,e-mail
address: dragovich@phy.bg.ac.yu}
\\ \normalsize{Institute of Physics, P.O. Box 57, 11001 Belgrade,}\\  \normalsize{Serbia and Montenegro}
\vspace{2mm} \\ \bf{Zoran Raki\' c}\hspace{.25mm}\thanks{\,e-mail
address: zrakic@matf.bg.ac.yu}
\\ \normalsize{Faculty of Mathematics, P.O. Box 550, 11001 Belgrade,}\\  \normalsize{ Serbia and Montenegro}}

\date{}

\maketitle


\begin{abstract}
We consider classical and quantum mechanics for an extended
Heisenberg algebra with additional canonical commutation relations
for position and momentum coordinates. In our approach this
additional noncommutativity is removed from the algebra by linear
transformation of coordinates and transmitted to the Hamiltonian
(Lagrangian). Since linear transformations do not change the
quadratic form of Hamiltonian (Lagrangian), and Feynman's path
integral has well-known exact expression for quadratic models, we
restricted our analysis to this class of physical systems. The
compact general formalism presented here can be easily realized in
any particular quadratic case. As an important example of
phenomenological interest, we explored model of a charged particle
in the noncommutative plane with perpendicular magnetic field. We
also introduced an effective Planck constant $\hbar_{eff}$ which
depends on  noncommutativity.
\end{abstract}

\bigskip

PACS numbers: 11.10.Nx, 03.65.Db, 03.65.-w

MSC2000: 81S40, 81T75,  81S10

Keywords: extended Heisenberg algebra, noncommutative quantum
mechanics, path integrals, quadratic Lagrangians
\bigskip

\section{ Introduction }

It is well known that the standard $n$-dimensional quantum
mechanics  (QM) is based on the Heisenberg algebra

\beq [\hat{x}_i ,\hat{p}_j ] = i\, \hbar\, \delta_{ij} , \quad
[\hat{x}_i ,\hat{x}_j ] = 0 , \quad    [\hat{p}_i ,\hat{p}_j ] = 0
,\quad i, j = 1, 2, \cdots, n , \label{1.1} \eeq

\noindent which  Hermitian operators for position and momentum
coordinates satisfy in the Hilbert space of any standard
quantum-mechanical system. However, in the last few years there
has been an intensive interest in investigation of noncommutative
QM (NCQM) which mainly starts from the commutation relations

\beq [\hat{x}_i ,\hat{p}_j ] = i\, \hbar\, \delta_{ij} , \quad
[\hat{x}_i ,\hat{x}_j ] = i\, \hbar \, \theta_{ij} , \quad
[\hat{p}_i ,\hat{p}_j ] = 0 ,\quad i, j = 1, 2, \cdots, n ,
\label{1.2} \eeq

\noindent where  $\theta_{ij}$ are constant elements of a real
antisymmetric $(\theta_{ij} =- \theta_{ji})$ $n \times n$-matrix
$\Theta$. Although the first considerations of noncommutativity
(NC) (\ref{1.2}) go back to the 1930's (see, e.g. \cite{szabo1})
the real excitement began in the 1998 when spatial NC of the form
$ [\hat{x}_i ,\hat{x}_j ] = i\, \hbar \, \theta_{ij}$ appeared in
the low energy string theory with D-branes  in a constant
background B-field (see reviews \cite{nekrasov}, \cite{szabo} and
references therein).

Adopting the NC (\ref{1.2}), one has wider than standard
uncertainty, i.e.

\beq \Delta x_i \, \Delta p_j \geq \frac{\hbar}{2}\,\delta_{ij} ,
\quad  \Delta x_i \,  \Delta x_j \geq
\frac{\hbar}{2}\,|\theta_{ij}| , \label{1.3} \eeq

 \noindent which prevents  from simultaneous accurate measuring not only
 usual $x_i$ and $p_i$ but also spatial coordinates
 $x_i$ and $x_j$  $\quad (i \neq j)$. One often takes $\theta_{ij} = \theta \,
 \varepsilon_{ij} ,$ where $(\varepsilon_{ij} ) = \mathcal E$ is the unit
 $n\times n$ antisymmetric matrix with $\varepsilon_{ij}= + 1 $ if $i < j$.
  Due to the uncertainty  $ \Delta x_i \,  \Delta x_j \geq
\frac{\hbar}{2}\,|\theta| ,  \quad (i \neq j), $ a spatial point
is  not a well  defined concept and the space becomes fuzzy at
distances of the order $\sqrt{\hbar\, |\theta|} ,$ which may be
much larger than the Planck or  string length.

 Most portion of the research in this subject has been mainly
 devoted to noncommutative
field theory (for reviews, see e.g. \cite{nekrasov} and
\cite{szabo}). NCQM has been also  actively explored. It is
motivated by the fact that NCQM can be regarded as the corresponding
one-particle nonrelativistic sector of noncommutative quantum field
theory. Also it provides construction of noncommutative models
suitable for the theoretical research and for potential experimental
tests. To this end, many two-dimensional models have been studied
and possible noncommutative contributions to the well-known effects
(Aharonov-Bohm effect \cite{chaichian}, lowest Landau level
\cite{szabo1}, fractional Hall effect \cite{dayi}, ...) have been
investigated.

Evolution in NCQM has been mainly investigated using the
Schr\"odinger equation. The path integral method has attracted
less attention, however for systems with quadratic Lagrangians a
systematic investigation started recently ( see
\cite{dragovich1}-\cite{dragovich3} and references therein).

 We consider here $n$-dimensional NCQM which is mainly based on the
following algebra

\bea\hp{-4mm} [ \hat{x_a},\hat{p_b}] = i\,\hbar\, ( \delta_{ab} -
\frac{1}{4}\, \theta_{ac}\, \sigma_{cb})\,, \quad
[\hat{x_a},\hat{x_b}] = i\,\hbar\, \theta_{ab}\,,\quad
[\hat{p_a},\hat{p_b}] =i\,\hbar\, \sigma_{ab}  \,, \label{1.4}
\eea

\noindent  where $(\theta_{ab}) = {\Theta}$ and $(\sigma_{ab})
={\Sigma}$ are the antisymmetric matrices with constant elements.
This kind of an extended noncommutativity maintains
 symmetry between canonical variables and yields (\ref{1.2}) in the limit
 $\sigma_{a b} \to 0$. We suppose that $|\theta \sigma |\ll 1$. The
algebra (\ref{1.4})  allows  simple reduction to the usual
commutation relations

\bea\hp{-5mm} [ \hat{q_a},\hat{k_b}] = i\,\hbar\, \delta_{ab},
\qquad\quad [\hat{q_a},\hat{q_b}] = 0,\qquad\quad
[\hat{k_a},\hat{k_b}] = 0 , \label{1.5} \eea

\noindent using the following linear transformations: \bea \hat{
x_a} = \hat{q_a} - \frac{\theta_{ab}\, \hat{k_b}}{2}\,,
\qquad\qquad \hat{ p_a} = \hat{k_a} + \frac{\sigma_{ab}\,
\hat{q_b}}{2}\, , \label{1.6} \eea

\noindent where summation over repeated indices is understood. It
was shown recently \cite{dragovich4} that NC (\ref{1.4}) is
suitable for study of possible dynamical control of decoherence by
applying perpendicular magnetic field to a charged particle in the
plane. This property also gives possibility to observe NC.

In this paper we present  compact general formalism of NCQM for
quadratic Lagrangians (Hamiltonians) with the above form of the
NC. The formalism developed is suitable for both Schr\"odinger and
Feynman approaches to quantum evolution.  There are now many
papers on some concrete models in NCQM, however, to our best
knowledge, there is no article on the evaluation of general
quadratic Lagrangians (Hamiltonians). Especially, Feynman's path
integral method to the NC has been almost ignored. Note that
quadratic Lagrangians contain an important class of physical
models, and that some of them are rather simple and exactly
solvable (a free particle, a particle in a constant field, a
harmonic oscillator). The obtained relations between coefficients
in commutative and noncommutative regimes give possibility to
easily construct effective Hamiltonians and Lagrangians in the
particular noncommutative cases.

Sec. 2 contains various expressions for quadratic Lagrangians and
Hamiltonians in commutative and noncommutative regimes, as well as
relations between them. In Sec. 3, the corresponding Schr\"odinger
equation and Feynman path integral are written down. As an example
of important phenomenological interest \cite{dragovich4}, a
charged particle in the noncommutative plane with homogeneous
perpendicular magnetic field is presented in Sec. 4. In the last
section, we discuss and emphasize some main results.

\bigskip

\section{Quadratic Lagrangians and Hamiltonians}

We start with general  quadratic Lagrangian for an $n$-dimensional
system, with position coordinates $x^T = (x_1 , x_2 , \cdots ,
x_n)$, which has the form \bea \hp{-7mm} L(\dot{x}, x,t) =
\frac{1}{2}\,\left( \dot{x}^T \alpha\, \dot{x} + \dot{x}^T \beta\,
x  + x^T \beta^T \dot{x} + x^T  \gamma\, x \right)+ \delta^T
\dot{x} + \eta^T x + \phi \,, \label{2.1} \eea

\noi where coefficients of the $n\times n$ matrices $\alpha
=((1+\delta_{ab})\, \alpha_{ab}(t)), $ \linebreak $ \beta
=(\beta_{ab}(t)),\ \gamma =((1+\delta_{ab} )\, \gamma_{ab}(t)),$
$n$-dimensional vectors $ \delta =(\delta_{a}(t)),$ \ $\eta
=(\eta_{a}(t))$ and a scalar $\phi =\phi(t)$ are some analytic
functions of  the time $t$. Matrices $\alpha$ and $\gamma$ are
symmetric,  $\alpha$ is nonsingular $(\det\alpha \neq 0)$ and
superscript ${}^T$ denotes transposition.

The  Lagrangian (\ref{2.1}) can be rewritten in the more compact
form: \bea L(X,t) = \frac{1}{2}\, X^T \, M\, X + N^T \, X + \phi ,
\label{2.2} \eea

\noi where $2n\times 2n$ matrix $M$ and $2n$-dimensional vectors
$X,\, N$ are defined as \bea \hp{-7mm}
 {M } = \left(
\begin{array}{ccc}
 \alpha & \beta  \\
 \beta^T & \gamma \end{array}
\right)\, , \qquad  X^T = (\dot{x}^T\, ,\,  x^T) \, , \qquad N^T =
(\delta^T\, ,\,  \eta^T) . \label{2.3} \eea

 Using the equations $ p_a = {\partial L \over \partial \dot
x_a},$ one finds $ {\dot x } = {\alpha^{-1}}\, ({p} - {\beta}\, {x
} - \delta ). $ Since the function $\dot{x}$ is linear in $p$ and
$x$, the corresponding classical Hamiltonian  $ H(p,x,t)= {p}^T\,
\dot {x } - L(\dot{x},x,t)$ becomes also quadratic, i.e.

\bea\hp{-9mm} H(p,x,t) = \frac{1}{2}\,\left( {p}^T \!\hp{-.2mm}
A\hp{.2mm} {p} + p^T\hp{-.5mm} B\hp{.2mm} x + x^T\hp{-.2mm}
 B^T\hp{-.2mm} p  +  x^T\hp{-.2mm}  C\hp{.2mm} x \right) + D^T\hp{-.2mm} p
 + E^T\hp{-.2mm} x + F
, \label{2.4} \eea

\noi where: \bea \hp{-8mm} \begin{array}{ll}  {A } =
{\alpha}^{-1}, \hspace{1.7cm} {B } =-\, {\alpha}^{-1}\, {\beta},
\hspace{1.7cm} {C} =
{\beta}^T\, {\alpha}^{-1}\, {\beta} - {\gamma } , \vp{1mm} \\
  {D} =- \, {\alpha }^{-1}\,  \delta, \hspace{1.0cm}  {E } =
\beta^T \, \alpha^{-1}\,  \delta - \eta , \hspace{1.0 cm} {F} =
\displaystyle{{1\over 2}}\, {\delta}^T \, {\alpha}^{-1} {\delta}\,
- {\phi } \, .
\end{array}  \label{2.5} \eea

Due to the symmetry of matrices $\alpha$ and $\gamma$ one can
easily see that matrices $\,A = ((1+\delta_{ab})\, A_{ab}(t))\,$
and $\,C = ((1+\delta_{ab})\, C_{ab}(t))\,$ are also symmetric
($A^T = A ,\, \, C^T = C$). The nonsingular $(\det {\alpha}\neq
0)$ Lagrangian $L(\dot{x},x,t)\, $   implies nonsingular $ (\det{
A}\neq 0) $ Hamiltonian $H(p,x,t) $. Note that the inverse map,
i.e. $H \to L$, is given by the same relations (11).

 The Hamiltonian (\ref{2.4}) can be also presented in the compact
form

\bea H(\Pi,t) = \frac{1}{2}\; \Pi^T  {\mathcal M}\; \Pi +
{\mathcal N}^T \,  \Pi   + F , \label{2.6} \eea

\noindent where matrix ${\mathcal M}$ and vectors $\Pi,\,
{\mathcal N}$ are

\bea \hp{-7mm}   {\mathcal M } = \left(
\begin{array}{ccc}
 A & B  \\
 B^T & C \end{array}
\right)\, , \qquad  \Pi^T = (p^T\, ,\,  x^T) \, , \qquad {\mathcal
N}^T = (D^T\, ,\,  E^T)\, . \label{2.7} \eea

\noi One can  show that

\bea  {\mathcal M}= \sum_{i =1}^3 \Upsilon_i^T(M)\,M\,
\Upsilon_i(M) , \label{2.8} \eea

\noi where \bea \hp{-7mm} \ba{l} \Upsilon_1(M)=\left( \ba{lr}
\al^{-1} &0\\0 & -I\ea\right), \qquad \qquad
\Upsilon_2(M)=\left( \ba{lr} 0\  &  \alpha^{-1} \beta\\
0 & 0\ea\right), \vp{4mm} \\
 \Upsilon_3(M)=\left( \ba{lr} 0\  &0\\0 & \
 i\sqrt{2}\,I\ea\right) ,\ea \label{2.9}
 \eea
 and $I$ is $n \times n$ unit matrix.
 One has also  ${\mathcal N}= Y(M) \, N,  $
 where

 \bea \hp{-7mm} Y (M)= \left(\ba{rr}
-\,\al^{-1} & 0  \\ \bet^T\,\al^{-1}& -I  \ea\right) = -\Upsilon_1
(M) + \Upsilon^T_2 (M) + i\, \sqrt{2}\,\, \Upsilon_3 (M)\,,
\label{2.10} \eea

\noi and $F = N^T\, Z(M)\, N - \phi ,$ where \bea \hp{-7mm} Z(M )
= \left(\ba{ll} \frac{1}{2}\, \al^{-1} & 0  \\ 0 & 0 \ea\right)
=\frac{1}{2}\, \Upsilon_1 (M) - \frac{i}{2\sqrt{2}}\, \Upsilon_3
(M)\, . \label{2.11}\eea

\noindent Using auxiliary matrices $\Upsilon_1 (M), \Upsilon_2
(M)$ and $\Upsilon_3 (M)$ in the above way, Hamiltonian quantities
${\mathcal M}, {\mathcal N}$ and $F$ are connected to the
corresponding Lagrangian ones $M, N$ and $\phi$.

 Eqs. (\ref{1.6}) can be rewritten in the compact form as

\bea\hp{-8mm} \hat{\Pi} = \Xi \,\, \hat{K} ,\quad \qquad \Xi =
\left(
\begin{array}{ccc}
 I & \frac12\,\, {\Sigma}  \\
- \frac12\,\, {\Theta} & I \end{array} \right) , \quad\qquad
\hat{K}= \left(
\begin{array}{ccc}
 \hat{k}  \\
 \hat{q} \end{array}
\right) .\  \label{2.12} \eea

\noi Since Hamiltonians depend on canonical variables, the
transformation  (\ref{2.12}) leads to the transformation of
Hamiltonians (\ref{2.4}) and (\ref{2.6}). To this end, let us
quantize the Hamiltonian (\ref{2.4}) and it easily becomes

\bea \hp{-8mm} H(\hat{p},\hat{x},t) = \frac{1}{2}\, (
{\hat{p}}^T\! A\, \hat{p} + \hat{p}^T \! B\, \hat{x} + \hat{x}^T
\! B^T\! \hat{p} + \hat{x}^T \! C\, \hat{x} ) + D^T\! \hat{p} +
E^T\! \hat{x} + F , \label{2.13} \eea because (\ref{2.4}) is
already written in the Weyl symmetric form.

Performing linear transformations (\ref{1.6}) in (\ref{2.13}) we
again obtain quadratic quantum Hamiltonian

\bea\nn \hp{-9mm} H_{\theta\sigma}(\hat{k},\hat{q},t) &=&
\frac{1}{2}\,\left( {\hat{k}}^T \hp{-.5mm} A_{\theta\sigma}\,
\hat{k} + \hat{k}^T \hp{-.5mm} B_{\theta\sigma}\, \hat{q} +
\hat{q}^T \hp{-.5mm} B^T_{\theta\sigma}\, \hat{k} + \hat{q}^T
\hp{-.5mm} C_{\theta\sigma}\, \hat{q} \right) \\ && \label{2.14}
 +\ D^T_{\theta\sigma}\, \hat{k} + E^T_{\theta\sigma}\, \hat{q} +
F_{\theta\sigma}\, , \label{2.14}   \eea where

\bea \hp{-8mm} \ba{ll} \displaystyle{{A}_{\theta\sigma} = {A} -
{1\over 2}\,\, {B}\, { \Theta} + {1\over 2}\,\, {\Theta}\, {B}^T
-{1\over 4}\,\, {\Theta}\, {C}\, { \Theta} ,} & \hp{5mm}
\displaystyle{{D}_{\theta\sigma} = {D} + {1\over
2}\,\, {\Theta}\, {E}}, \vp{2mm} \\
\displaystyle{{B}_{\theta\sigma} = {B} + \frac{1}{2}\, {\Theta}\,
{C} + \frac{1}{2}\, A\, \Sigma + \frac14\,\, \Theta\, B^T \,
\Sigma} , & \hp{5mm} \displaystyle{{E}_{\theta\sigma} = {E}-
\frac 12\, \Sigma\, D} ,  \vp{2mm}\\
\displaystyle{{C}_{\theta\sigma} = {C} - {1\over 2}\,\Sigma\, {B}
+ {1\over 2}\, B^T \, {\Sigma} -{1\over 4}\, {\Sigma}\, {A}\,
\Sigma}, & \hp{5mm} \displaystyle{{F}_{\theta\sigma} = {F}} .\ea
 \label{2.15}
\eea Note that for the nonsingular Hamiltonian $
H(\hat{p},\hat{x},t)$ and for sufficiently small $\theta_{ab}$ the
Hamiltonian $ H_{\theta\sigma}(\hat{k},\hat{q},t)$ is also
nonsingular.  $A_{\theta\sigma}$ and $D_{\theta\sigma}$ do not
depend on $\sigma$, as well as $C_{\theta\sigma}$ and
$E_{\theta\sigma}$ do not contain $\theta$.

It is worth noting that Hamiltonian (\ref{2.14}) with
noncommutativity (\ref{1.4}) is equivalent to Hamiltonian
(\ref{2.15}) with ordinary commutation relations (\ref{1.5}).
Dynamics  depends on parameters $\theta$ and $\sigma$, what is more
obvious from Hamiltonian (\ref{2.14}) than relations (\ref{1.4}).
Consequently phase space transformations (\ref{1.6}) are not (and
should not be) canonical, because dynamics of our system is given
not only by Hamiltonian (\ref{2.14}) but with (\ref{2.14}) and
(\ref{1.4}). In the case $\sigma = 0$, a similar result for
Hamiltonian with parameter $\theta$ is obtained by the Moyal product
procedure \cite{mezincescu}.

Classical analogue of (20) maintains the same form

\bea\nn \hp{-7mm} H_{\theta\sigma}(k,q,t) &=& \frac{1}{2}\, (
k^T\hp{-.5mm} A_{\theta\sigma}\, {k} + {k}^T \hp{-.5mm}
B_{\theta\sigma}\, {q} + {q}^T\hp{-.2mm} B^T_{\theta\sigma}\, {k}
+ {q}^T\hp{-.2mm} C_{\theta\sigma}\, {q} ) \\ && \nn  +\
D^T_{\theta\sigma}\, {k} + E^T_{\theta\sigma}\, {q} +
F_{\theta\sigma}\,.\eea

\noi In the compact form Hamiltonian (\ref{2.14})  becomes

\bea\hp{-7mm}
 \hat
H_{\theta\sigma}(\hat{K},t) = \frac{1}{2}\,\, \hat K^T \,
{\mathcal M}_{\theta\sigma}\; \hat K + {\mathcal
N}_{\theta\sigma}^T \; \hat K + F_{\theta\sigma} , \label{2.16}
\eea

\noi where $2n\times 2n$ matrix ${\mathcal M}_{\theta\sigma}$ and
$2n$-dimensional vectors $ {\mathcal N}_{\theta\sigma}, \,
\hat{K}\,$ are

\bea\hp{-9mm}  {\mathcal M}_{\theta\sigma} = \left(
\begin{array}{ccc}
 A_{\theta\sigma} & B_{\theta\sigma}  \\
 B_{\theta\sigma}^T & C_{\theta\sigma} \end{array}
\right)   , \quad\ \ {\mathcal N}_{\theta\sigma}^T =
(D_{\theta\sigma}^T\,, \, E_{\theta\sigma}^T) , \ \ \quad
\hat{K}^T = (\hat{k}^T\,, \,  \hat{q}^T)
 . \label{2.17} \eea

 \noi From (\ref{2.6}), (\ref{2.12}) and (\ref{2.16}) one can find connections between
${\mathcal M}_{\theta\sigma},\, {\mathcal N}_{\theta\sigma} , \, {
F}_{\theta\sigma} $ and ${\mathcal M},\, {\mathcal N}, \, F$,
which are given by the following relations:

\bea\hp{-10mm} {\mathcal M}_{\theta\sigma}=\Xi^T\,{\mathcal M}\,\,
\Xi \,, \quad\qquad {\mathcal N}_{\theta\sigma}=\Xi^T\,{\mathcal
N} , \qquad\quad {F}_{\theta\sigma} = F.  \label{2.18}  \eea

Using equations $ \dot{q}_a = \frac{\partial
H_{\theta\sigma}}{\partial k_a} $ which give $ k =
A_{\theta\sigma}^{-1}\, (\dot{q} - B_{\theta\sigma}\, q -
D_{\theta\sigma}), $ we can pass from the classical form of
Hamiltonian (20) to the corresponding Lagrangian by relation $
L_{\theta\sigma} (\dot{q},q,t) = k^T \dot{q}  - H_{\theta\sigma}
(k,q,t) . $ Note that coordinates $q_a$ and $x_a$ coincide when
$\theta =\sigma =0$. Performing necessary computations we obtain

\bea \nn \hp{-8mm} L_{\theta\sigma}(\dot{q}, q,t) &=&
\frac{1}{2}\,\left( \dot{q}^T  \alpha_{\theta\sigma}\, \dot{q} +
\dot{q}^T \beta_{\theta\sigma}\, q  +  q^T
\beta^T_{\theta\sigma}\,
\dot{q}  +  q^T  \gamma_{\theta\sigma}\, q \right)  \\
 && +\  \delta^T_{\theta\sigma}\, \dot{q} + \eta^T_{\theta\sigma}\, q
+ \phi_{\theta\sigma} , \label{2.19} \eea

\noi or in the  compact form: \bea\hp{-8mm} L_{\theta\sigma}(Q,t)
= \frac{1}{2}\; Q^T \, M_{\theta\sigma}\; Q+ N_{\theta\sigma}^T \;
Q + \phi_{\theta\sigma} , \label{2.20} \eea

\noi where \bea \hp{-7mm} {M_{\theta\sigma}} = \left(
\begin{array}{ccc}
 \alpha_{\theta\sigma} & \beta_{\theta\sigma}  \\
 \beta_{\theta\sigma}^T & \gamma_{\theta\sigma} \end{array}
\right)\, ,  \quad N_{\theta\sigma}^T =
(\delta_{\theta\sigma}^T\,, \, \eta_{\theta\sigma}^T)\, , \quad
Q^T = (\dot{q}^T\,, \,  q^T) \, , . \label{2.21} \eea

\noi Then the connections between ${M_{\theta\sigma}},\,
{N_{\theta\sigma}},\, \phi_{\theta\sigma} \, $ and $\, M,\,
N,\,\phi$ are given by the following relations:

\bea \hp{-9mm} \ba{l} M_{\theta\sigma} = \sum\limits_{i,j=1}^3 \,
\Xi_{ij}^T\, \, M\, \,\Xi_{ij},\qquad \qquad\hp{1.6mm}
  \Xi_{ij}=\Upsilon_i(M)\,\,\Xi\,\,\Upsilon_j(\mathcal{M}_{\theta\sigma})
, \vp{3mm} \\
 N_{\theta \sigma} = Y (\mathcal{M}_{\theta\sigma})\, \Xi^T\,
Y(M)\, N, \quad \quad \phi_{\theta \sigma} =
\mathcal{N}_{\theta\sigma}^T \, Z (\mathcal{M}_{\theta\sigma}) \,
\mathcal{N}_{\theta\sigma} - F  \, .\ea  \label{2.22} \eea

 \noi In more detail, the connection between  coefficients of the
 Lagrangians $L_{\theta\sigma}$
 and  $L$ is given by the relations:

 \bea \hp{-7mm}\ba{l} {\al}_{\theta\sigma}  = \big[\,
{\al}^{-\,1} - {\frac 12} \, (\Theta\, {\bet}^{T}\, {\al}^{-\,1}-
{\al}^{-\,1}\, {\bet}\, \Theta) -\frac14\,\Theta\,( {\bet}^{T}\,
{\al }^{-\,1}\, {\bet} - \gam)
\,\Theta \, \big]^{-\,1}\,, \vp{2mm}  \\
  {\bet}_{\theta\sigma}  =  {\al}_{\theta\sigma}\,
 \big( {\al}^{-\,1}\,
 {\bet}  - {\frac 12}\,( {\al}^{-\,1}\,\Sigma -\Theta\, {\gam}+
  \Theta\, {\bet}^{T}\, {\al}^{-\,1}\,{\bet})
  +\frac14\,\Theta\,{\bet}^{T}\,{\al}^{-\,1}\,\Sigma\big)\,, \vp{2mm} \\
 {\gam}_{\theta\sigma} = \gamma +  \bet_{\theta\sigma}^T \,
\al_{\theta\sigma}^{-\,1} \, \bet_{\theta\sigma} -  {\bet }^{T}\,
{\al}^{-\,1}\, {\bet}  + {\frac 14}\, \Sigma\,{\al}^{-\,1}\,
\Sigma\,\vp{1mm} \\ \hp{11mm}  -\ {\frac
12}\,(\Sigma\,{\al}^{-\,1}\,\bet -{\bet }^{T}\,
{\al}^{-\,1}\,\Sigma ) \,, \vp{2mm} \\
 {\delta}_{\theta\sigma}  =
{\al}_{\theta\sigma}\,\big({\al}^{-\,1}\,{\delta}+\frac
12\,\,(\Theta\,\eta- \Theta\,{\bet}^{T}\,
{\al}^{-\,1}\,{\delta}) \big)\,, \vp{2mm}  \\
 {\eta }_{\theta\sigma} = \eta + \bet_{\theta\sigma}^T \,
\al_{\theta\sigma}^{-\,1} \, \delta_{\theta\sigma} -
{\bet}^{T}\,{\al}^{-\,1}\, {\delta}  - \frac 12\,\,
\Sigma\,{\al}^{-\,1}\,{\delta}\,,\vp{2mm}  \\
 {\phi }_{\theta\sigma}  =  \phi + \frac 12\,\,
{\delta}_{\theta\sigma}^T\,  \al_{\theta\sigma}^{-\,1} \,
\delta_{\theta\sigma} - \frac 12\,\, \delta^T\,
{\al}^{-\,1}\,{\delta} \,. \ea
 \label{2.23} \eea
Note that $\alpha_{\theta \sigma}, \,  \delta_{\theta \sigma} $
and $ \phi_{\theta \sigma}$ do not depend on $\sigma$. \vp{3mm}

\bigskip

\section{Noncommutative Schr\"odinger Equation and Path Integral}

The corresponding Schr\"odinger equation in this NCQM is

\bea\hp{-4mm} i\, \hbar \frac{\partial \Psi (q, t)}{\partial t} =
{{H}_{\theta\sigma} (\hat{k}, q, t)}\, \Psi (q, t) \,, \label{3.1}
\eea

\noi where $ \hat{k}_a = -\, i\, \hbar \frac{\partial}{\partial
q_a} , \, \, a = 1,2, \cdots, n$ and $ {H}_{\theta\sigma}
(\hat{k}, q, t) $ is given by (\ref{2.14}). Investigations of
dynamical evolution have been mainly performed using the
Schr\"odinger equation and this aspect of NCQM is much more
developed than  the noncommutative Feynman path integral. For this
reason and importance of  the path integral method, we will in the
sequel give priority  to the description of this approach.

To compute a path integral, which is a basic instrument in
Feynman's approach to quantum mechanics, one can start from its
Hamiltonian formulation on the phase space. However, when
Hamiltonian is a quadratic polynomial with respect to momentum $k$
(see, e.g. \cite{dragovich2}) such path integral on a phase space
can be reduced to the Lagrangian path integral on configuration
space. Hence, for the Hamiltonians (\ref{2.14}) and (\ref{2.16})
we have derived the corresponding Lagrangians (\ref{2.19}) and
(\ref{2.20}).

 The  standard Feynman path integral \cite{feynman} is
  \bea
\hp{-4mm} {\mathcal K} (x'',t'';x',t') =\int_{x'}^{x''} \exp \left
( \frac{i}{\hbar}\, \int_{t'}^{t''} L(\dot{q},q, t)\, dt \right
)\, {\mathcal D}q \,, \label{3.2} \eea

\noi where ${\mathcal K}(x'',t'';x',t')$ is the kernel of the
unitary evolution operator $U (t)$ and $x''=q(t''), \ x'=q(t')$
are end points.
 In ordinary quantum mechanics (OQM), Feynman's path integral for quadratic
 Lagrangians  can be
evaluated analytically and its exact  expression has the form
\cite{steiner} \bea
 \hp{-6mm} {\mathcal K}(x'',t'';x',t') =\frac{1}{(i
h)^{\frac{n}{2}}} \sqrt{\det{\left(-\frac{\partial^2 {\bar
S}}{\partial x''_a
\partial x'_b} \right)}} \exp \left(\frac{2\pi i}{h}\,{\bar
S}(x'',t'';x',t')\right), \label{3.3} \eea

\noi where $ {\bar S}(x'',t'';x',t')$ is the action for the
classical trajectory. According to (\ref{1.4}), (\ref{1.5}) and
(\ref{1.6}), NCQM related to the quantum phase space $(\hat{p} ,\,
\hat{x})$ can be regarded as an OQM on the standard phase space
$(\hat{k} ,\, \hat{q})$ and one can apply usual path integral
formalism.

A systematic path integral approach to NCQM with quadratic
Lagrangians (Hamiltonians) has been investigated during the last
few years in \cite{dragovich1}, \cite{dragovich2} and
\cite{dragovich3}. In \cite{dragovich1} and \cite{dragovich2},
general connections between quadratic Lagrangians and Hamiltonians
for standard and $\theta \neq 0$, $\sigma =0$ NC are established,
and this formalism was applied to a particle in the
two-dimensional noncommutative plane with a constant field and to
the noncommutative harmonic oscillator. Paper \cite{dragovich3}
presents generalization of articles \cite{dragovich1} and
\cite{dragovich2} towards noncommutativity (\ref{1.4}). The
present article develops formalism of \cite{dragovich3} and
contains its application to a charged particle in the
noncommutative plane with homogeneous electric and perpendicular
magnetic field.

If we know Lagrangian (\ref{2.1}) and algebra (\ref{1.4}) we can
obtain the corresponding effective Lagrangian (\ref{2.19})
suitable for the path integral in NCQM. Exploiting the
Euler-Lagrange equations \bea\nn \frac{\partial
L_{\theta\sigma}}{\partial q_a} -\frac{d}{dt} \frac{\partial
L_{\theta\sigma}} {\partial{\dot q}_a} =0 \,, \quad a = 1, 2,
\cdots, n \,, \eea one can obtain the classical trajectory $q_a
=q_a (t)$ connecting end points $x' = q(t')$ and $x''= q(t'')$,
and the corresponding action is
 \bea\nn {\bar S}_{\theta\sigma}
(x'',t'';x',t') =\int_{t'}^{t''} L_{\theta\sigma} (\dot{q}, q,t)\,
dt \,.\eea Path integral in NCQM is a direct analogue of
(\ref{3.3}) and its exact expression in the form of quadratic
actions ${\bar S_{\theta\sigma}}(x'',t'';x',t')$ is \bea &&
{\mathcal K}_{\theta\sigma}(x'',t'';x',t') = \frac{1}{(i
h)^{\frac{n}{2}}} \sqrt{\det{\left(-\frac{\partial^2 {\bar
S_{\theta\sigma}}}{\partial x''_a\,
\partial x'_b} \right)}}  \exp \left(\frac{2\,\pi\, i}{h}\,{\bar
S_{\theta\sigma}}(x'',t'';x',t')\right) .\nn \\ &&  \label{3.4}
\eea \vp{3mm}

\bigskip

 \section{Particle in a Noncommutative Plane with Electric
 and Perpendicular Magnetic Field }

To illustrate many features of the above formalism, as well as to
give a model suitable for potential phenomenology (see, e.g.
\cite{dragovich4}), we consider a particle of a charge $e $ moving
in a plane $(x_1, x_2 )$ with noncommutativity parameters $\theta
$ and $\sigma$. Let this particle be also under the influence of a
homogeneous electric field ${\mathcal E}$ along coordinate $x_1$
and a homogeneous magnetic field ${\mathcal B}$ perpendicular to
the plane and oriented along the axis $-x_3$. It is convenient to
start from the nonrelativistic Hamiltonian \bea H(p, x) =
\frac{1}{2\, m} \, \Big[ (p_1 - e\, {\mathcal A}_1)^2 + (p_2 - e\,
{\mathcal A}_2)^2 \Big] + e \, \varphi , \label{4.1} \eea

\noi where    ${\mathcal A}_1 =\frac{{\mathcal B} }{2}\, x_2\, ,
\, \, {\mathcal A}_2 =-\,\frac{{\mathcal B} }{2}\, x_1$ and
$\varphi = - {\mathcal E}\, x_1$ . This dynamical system is also
very suitable to study possible phenomenological aspects of NCQM.
Using the inverse map of (\ref{2.5}) we get the corresponding
Lagrangian \bea\hp{-5mm}
 L (\dot{x}, x) = \frac{m}{2}\, (\dot{x}_1^2 +
\dot{x}_2^2 )  + \frac{e\, {\mathcal B}}{2}\, (\dot{x}_1 x_2 -
\dot{x}_2 x_1) + e \hp{.7mm} {\mathcal E}\hp{.3mm} x_1 .
\label{4.2} \eea

\noi Employing formulas  (\ref{2.15}) and (\ref{2.23}),  one
obtains respectively Hamiltonian \bea H_{\theta\sigma} (k, q) &=&
\frac{1}{2\hp{.3mm} \mu}\, (k_1^2 + k_2^2) -
\frac{\lambda}{2\hp{.3mm} \mu}\, (k_1\hp{.3mm} q_2 -
k_2\hp{.5mm} q_1) + \frac{\lambda^2}{8\hp{.3mm} \mu}\, (q_1^2  + q_2^2) \nn \\
&&  +\ \frac{\theta\hp{.3mm} e\hp{.6mm} {\mathcal E}}{2}\, k_2 -
e\hp{.6mm} {\mathcal E}\hp{.2mm} q_1 \label{4.3} \eea

\noi and Lagrangian \bea L_{\theta\sigma} (\dot{q}, q) =
\frac{\mu}{2} \, (\dot{q}_1^2 + \dot{q}_2^2) + \frac{\lambda}{2}
\, (\dot{q}_1\hp{.3mm} q_2 - \dot{q}_2\, q_1)  -
\frac{\mu\hp{.3mm} \theta\hp{.2mm} e\hp{.6mm} {\mathcal E}}{2}  \,
\dot{q}_2 + \nu_0\, q_1 + \frac{\mu\hp{.5mm} \theta^2\hp{.2mm}
e^2\hp{.5mm} {\mathcal E}^2}{8} \, , \label{4.4} \eea

\noi where
 \bea \hp{-4mm}
 \mu = \frac{m}{\big( 1- \frac{\theta\, e\,
{\mathcal B}}{4} \big)^2}\, \, , \quad \quad \lambda = \frac{e\,
{\mathcal B} - \sigma}{ 1- \frac{\theta\, e\, {\mathcal
B}}{4}}\,\,, \qquad \nu_0=e\, {\mathcal{E}}\,\Big(1+\frac{\theta\,
\lambda}{4}\Big)\,.   \label{4.5}\eea
 The
above Hamiltonian and Lagrangian are related to the dynamics in
noncommutative plane, where NC is characterized by parameters
$\theta$ and $\sigma$.

The Lagrangian given by (\ref{4.4}) implies the Euler-Lagrange
equations, \bea \hp{-5mm} \displaystyle{\mu\,\ddot{q}_1+
\lambda\,\dot{q}_2 =\nu_0,} \qquad \qquad
\displaystyle{\mu\,\ddot{q}_2- \lambda\,\dot{q}_1 =0\,.}
\label{4.6} \eea

\noi We transform the system (\ref{4.6})  to two uncoupled third
order differential equations \bea \ba{l}
\displaystyle{\mu^2\,{q}_1^{(3)}+ \lambda^2\, {q}_1^{(1)}=0,}
\hspace{15mm} \displaystyle{\mu^2\, {q}_2^{\,(3)}+
\lambda^2\,{q}_2^{(1)}-\lambda\, \nu_0=0\,.} \ea \label{4.7} \eea

\noi The solution of the equations (\ref{4.7}) has the following
form \bea \hspace{-16mm}  \ba{ll}  {q}_1(t) = C_1 + C_2 \cos
(\eta\, t) + C_3 \sin (\eta\, t)\,,   \vp{2mm}\\
{q}_2(t) = D_1 + D_2\, \cos (\eta\, t) + D_3 \sin (\eta\, t)+
\dst{\frac{\nu_0}{\lambda}}\,t\,, \qquad \eta=
\dst{\frac{\lambda}{\mu}}\,. \ea \label{4.8} \eea

\noi Imposing connection between $q_1$ and $q_2$ by  coupled
differential equations (\ref{4.6}), we obtain  $ C_2=D_3 $ and $
D_2=-C_3\,. $ The  constants $C_1,D_1,C_3$ and $D_3$  can be fixed
from initial conditions given by \bea \hp{-10mm}
q_1(0)=x_1',\qquad q_1(T)=x_1'', \qquad q_2(0)=x_2',\qquad
q_2(T)=x_2''.  \label{4.9} \eea

\noi By this way the corresponding constants become : \bea
 \hp{-10mm} \ba{l} \displaystyle{C_1=
\frac{x_1'+x_1''}{2}+
\frac{x_2'-x_2''}{2}\,\cot\!\Big[\hp{.2mm}\frac{T\hp{.3mm}\eta}{2}\hp{.2mm}\Big]+
\frac{T\hp{.3mm}\nu_0}{2\hp{.3mm}\lambda}\,
\cot\!\Big[\hp{.2mm}\frac{T\,\eta}{2}\hp{.2mm}\Big], } \vp{2mm}\\
  \displaystyle{C_3=
\frac{-x_2'+x_2''}{2}-
\frac{x_1'-x_1''}{2}\,\cot\!\Big[\hp{.2mm}\frac{T\hp{.3mm}\eta}{2}\hp{.2mm}\Big]-
\frac{T\hp{.3mm}\nu_0}{2\hp{.3mm}\lambda}\,, }
  \vspace{2mm} \\
\displaystyle{D_1= \frac{x_2'+x_2''}{2}-
\frac{x_1'-x_1''}{2}\,\cot\!\Big[\hp{.2mm}\frac{T\hp{.3mm}\eta}{2}\hp{.2mm}\Big]-
\frac{T\hp{.3mm}\nu_0}{2\,\lambda}\,, }
  \vspace{2mm} \\
  \displaystyle{D_3=
\frac{x_1'-x_1''}{2}-
\frac{x_2'-x_2''}{2}\,\cot\!\Big[\hp{.2mm}\frac{T\hp{.3mm}\eta}{2}\hp{.2mm}\Big]-
\frac{T\hp{.3mm}\nu_0}{2\hp{.3mm}\lambda}\,
\cot\!\Big[\hp{.2mm}\frac{T\hp{.3mm}\eta}{2}\hp{.2mm}\Big] \,.}
\ea \label{4.10} \eea

\noi Inserting the above expressions for constants (\ref{4.10})
into
 (\ref{4.8}) we obtain solutions of the Euler--Lagrange equations (\ref{4.6}).
 For these solutions and their time derivatives  we
find the following expression for Lagrangian (\ref{4.4}):
\bea\hp{-7mm} \ba{l} L_{\theta\sigma}(\dot q,q) =
\dst{\frac{1}{8\hp{.3mm}\lambda^2\hp{.3mm}\mu}}\Big(\mu\big(
4\hp{.3mm}C_1\hp{.3mm}\lambda^2 \hp{.3mm} \nu_0
+\mu\hp{.3mm}(-2\hp{.3mm}\nu_0\ +\theta\hp{.3mm}
\lambda\hp{.3mm} e\hp{.7mm} {\mathcal{E}})^2\big)+4\hp{.3mm} \lambda^2 \vp{1.0mm} \\
\hp{3mm} \times \, \Big( \big(
C_3\hp{.3mm}\lambda\hp{.3mm}(D_1\hp{.3mm}\lambda+\nu_0\hp{.3mm}t)
 - D_3\hp{.3mm}(C_1\hp{.3mm}\lambda^2 -3\hp{.4mm}\mu\hp{.3mm}\nu_0  +
\theta\hp{.3mm}\lambda\hp{.4mm}\mu\hp{.3mm}
e\hp{.7mm}{\mathcal{E}})\big)\hp{.3mm}\cos[\hp{.2mm}\eta\hp{.3mm}t\hp{.2mm}]
\vp{1.0mm}\\
 \hp{3mm} -\, \big( (C_1\hp{.3mm}C_3 + D_1\hp{.3mm}D_3)\hp{.3mm} \lambda^2
 + D_3\hp{.3mm}\lambda\hp{.3mm}\nu_0\hp{.3mm}t -3\hp{.4mm}C_3\hp{.3mm}\mu\hp{.3mm}\nu_0 + C_3\hp{.3mm}
\theta\hp{.3mm}\lambda\hp{.4mm}\mu\hp{.3mm}
e\hp{.7mm}{\mathcal{E}}\big)
 \hp{.3mm}\sin[\hp{.2mm}\eta\hp{.3mm}t\hp{.2mm}]\Big)\!\Big)\,. \ea \label{4.11}\eea

 \noi Using (\ref{4.11}), we finally compute the
corresponding classical action \bea \label{4.12}  \hp{-7mm} \ba{l}
{\bar S}_{\theta\sigma} (x'',T;x',0) =
\hp{-.4mm}\dst{\int\limits_{0}^T \! L_{\theta\sigma}(\dot{q},q)\,
d\hspace{.1mm}t} =
 \dst{\frac{\lambda}{2}\hp{-.2mm}\Big(\hp{-.2mm}x'_2\,x''_1-x'_1\,x''_2\hp{-.2mm} \Big)+
 \frac{\nu_0\hp{.2mm} T}{2}\hp{-.2mm}\Big(\hp{-.2mm}x'_1+x''_1\hp{-.2mm}\Big)} \vp{1.5mm} \\
\hp{7mm}  \dst{+\
\frac{\mu\hp{.2mm}\nu_0}{\lambda}\hp{-.2mm}\Big(\hp{-.2mm}x''_2-x'_2\hp{-.2mm}\Big)+\frac{\theta\hp{.2mm}\mu\hp{.2mm}
e\hp{.6mm}
{\mathcal{E}}}{2}\hp{-.2mm}\Big(\hp{-.2mm}x'_2-x''_2\hp{-.2mm}\Big)+
  \frac{\mu\hp{.2mm}T}{8}\hp{-.2mm}\Big(\hp{-.2mm}-4\,\frac{\nu_0^2}{\lambda^2}+
 \theta^2\hp{.2mm}e^2\hp{.3mm}{\mathcal{E}}^2\hp{-.2mm}\Big)} \vp{2.0mm}  \\
 \hp{7mm} \dst{+\ \frac{(x'_1-x''_1)^2\,\lambda^2 + ((x'_2-x''_2)\,\lambda+T\,
\nu_0)^2}{4\,\lambda} \,
\cot\!\Big[\hp{.2mm}\frac{\eta\,t}{2}\hp{.2mm}\Big] \, .} \ea \eea

\noi Finally, we obtain the determinant  \bea \hp{-8mm}
\det{\left(\!-\frac{\partial^2 {\bar S_{\theta\sigma}}}{\partial
x''_a\,
\partial x'_b}\right)} =\frac{\lambda^2}{4\,
\sin^2\big[\frac{\lambda\,T}{2}\big]}\,\,, \label{4.13}\eea

\noi and the transition probability amplitude to reach the point
$(x'' , T)$ from the point $(x' , 0)$ : \bea\hp{-5mm}
  {\cal K}_{\theta\sigma} (x'',T;x',0) =
 \frac{|\hp{.4mm}\lambda\,\big|}
 {2\,i\, h\,\big|\!\sin\big[\frac{\lambda\,T}{2}\big]\hp{-.1mm}\big|}\,
 \exp\left(
 \frac{2\pi i}{h}\, \bar{S}_{\theta\sigma} (x'',T;x',0)   \right)
,  \label{4.14} \eea

\noi where $\bar{S}_{\theta\sigma} (x'',T;x',0)$ is given by
(\ref{4.12}). According to the Feynman approach to QM, all
information on our noncommutative quantum system is encoded in the
$ {\cal K}_{\theta\sigma} (x'',T;x',0)$ in (\ref{4.14}).

\bigskip

\section{Discussion and Concluding Remarks}

 Let us mention  that taking $\,\sigma=0 ,\, \,  \theta=0\,$ in
the above formulas we recover expressions for the Lagrangian
$\,L(\dot{q}, q),\,$ classical action $\,\bar{S} (x'',T;x',0)\, $
and probability amplitude $\,{\mathcal K} (x'',T;x',0)\,$  of the
ordinary commutative case.

 Note that a similar path integral approach with $\,\sigma =0\,$
 has been considered
in the context of the Aharonov-Bohm effect \cite{chaichian}, the
Casimir effect, a quantum system in a rotating frame
\cite{christiansen}, and the Hall effect \cite{dayi} ( see also
reherences in \cite{dragovich1,dragovich2,dragovich3}). Our
investigation contains all quantum-mechanical systems with quadratic
Hamiltonians (\ref{2.13}) (Lagrangians (\ref{2.1})) on
noncommutative phase space given by (\ref{1.4}).

 On the basis of the expressions presented in this article, there
are many possibilities to discuss noncommutative
quantum-mechanical systems with respect to various values of
noncommutativity parameters $\,\theta\,$ and $\,\sigma.\,$ We will
discuss only some aspects of the two-dimensional model from the
preceding section taking electric field $\mathcal{E} = 0$.

 Since \bea \hp{-9mm} [\hp{.2mm}\hat{\pi}_a ,
\hat{\pi}_b\hp{.2mm}] = -\hp{.2mm} i\hp{.2mm} \hbar\, ( e\hp{.3mm}
{\mathcal B} - \sigma) \Big(1 - \frac{e\hp{.3mm} {\mathcal B}
\hp{.3mm}\theta}{4} \Big)\, \varepsilon_{ab} \,,      \label{5.1}
\eea

\noi where $ \, \hat{\pi}_a = \hat{p}_a -e \hp{.2mm}\hat{\mathcal
A}_a, $ a charged particle in the plane with the homogeneous
perpendicular magnetic field $\,{\mathcal B}\,$ and phase space NC
(\ref{1.4}) has motion under an effective momentum NC depending on
 $\,  -\hp{.2mm} ( e\hp{.3mm} {\mathcal B} - \sigma)
 \big(1 - \frac{e\hp{.3mm} {\mathcal B}\hp{.2mm} \theta}{4}
\big)$. Recall that $\,p\,$ is a canonical momentum and $\,\pi\,$
is the corresponding physical one.

Note that the NC parameter $\sigma$ appears in the form $
e{\mathcal B} - \sigma $, i.e. $\sigma$ and $e \, {\mathcal B}$
are on the equal footing. Accordingly, magnetic field ${\mathcal
B}$ can be regarded as a generator of the momentum NC with $\sigma
= - e {\mathcal B}$. In our galaxy and other galaxies, there is
evidence (see, e.g. \cite{bamba} and \cite{widrow}) of magnetic
field ${\mathcal B} \sim 10^{-10} T$, $\quad (T = Tesla)$ which
may generate  an effective $\sigma\sim 10^{-29} C T ,\, \, (C =
Coulomb), $ relevant to dynamics of electrons and protons. In the
expression $ e {\mathcal B} - \sigma $ one can regard $\sigma$ as
a result of an effective background magnetic field  and $e \,
\mathcal B$ as the effect of a laboratory magnetic field.

It is also worth noting that there is a sense to introduce an
effective reduced Planck constant as \bea \hbar_{eff} = \hbar \,
\Big( 1 + \frac{\theta \sigma}{4} \Big)\,, \label{5.2} \eea

\noi where one can take $\sigma = - e {\mathcal B} .$ The present
values of $\theta $ and $\sigma$ are very small so that
$\frac{\theta \sigma}{4} \ll 1$ and consequently
$\hbar_{eff}\approx \hbar$, but at very early times of the
universe evolution the situation might be very different. Suppose
that at least one of parameters $\theta$ and $\sigma$ depends on
cosmic time and let at very early stage of the evolution
$\frac{\theta \sigma}{4} \gg 1$ and after that it decreased to a
small value rapidly. Then $\frac{\theta \sigma}{4} \gg 1$ imposes
$\hbar_{eff} \gg \hbar$ and the universe was in a strengthened
quantum state, which could significantly change our view of its
beginning.  It follows that existence of $\theta\neq 0$ and
$\sigma\neq 0$, where the corresponding $\sigma$ can be generated
by magnetic field, may modify $\hbar$, and this should play also
some role at very high energies.

 Using annihilation and creation operators \bea\hp{-10mm} a =
\frac{\hat{\pi_1} - i\hp{.2mm}
\hat{\pi_2}}{\sqrt{\hp{.2mm}2\hp{.3mm} \hbar_{eff}\hp{.3mm}
m\hp{.3mm} {\omega}}}\, , \qquad\qquad a^\dagger  =
\frac{\hat{\pi_1} + i\hp{.2mm}
\hat{\pi_2}}{\sqrt{\hp{.2mm}2\hp{.3mm} \hbar_{eff}\hp{.3mm} m
\hp{.3mm}{\omega}}}\,, \label{5.3} \eea

\noi where the frequency \bea \hp{-9mm}{\omega} = \frac{
e\hp{.3mm} {\mathcal B} - \sigma}{m}\,, \label{5.4} \eea

\noi one can write the Hamiltonian (\ref{4.3}), with
noncommutative phase space (\ref{1.4}) and the electric field
$\,{\mathcal E} = 0\,,$ in the harmonic oscillator form \bea
\hp{-10mm}\hat{H} = {\omega}\hp{.3mm} \hbar_{eff} \hp{.3mm}
\Big(\hp{.3mm} a^\dagger\hp{.3mm} a  + \frac{1}{2} \hp{.3mm} \Big)
\label{5.5} \eea

\noi with energy levels $\,E_n = \omega\hp{.3mm} \hbar_{eff}
\hp{.3mm} (n + \frac{1}{2})\,, $ where now $\hbar_{eff} = \hbar \,
\big( 1 - \frac{e\, {\mathcal B}\, \theta}{4} \big)$. This is an
extended Landau problem which reduces to the standard one if
$\,\sigma = \theta = 0.\,$

\noi The frequency $\,\omega\,$ can be at will controlled  by
magnetic field $\,{\mathcal B}.\,$ In particular, $\,\omega = 0\,$
if $\,{\mathcal B} =\frac{\sigma}{e}\,,$ and  $\,\omega_{{\mathcal
B} \to 0} = - \frac{\sigma}{m}\,$ as well as $\,\omega =
\frac{e\hp{.2mm} {\mathcal B}}{ m}\,$ if $e {\mathcal B} \gg
\sigma$.

 In the very strong magnetic field $\,{\mathcal B}\,$ and $\,{\mathcal E} = 0\,,$
  the Lagrangian (\ref{4.4})  becomes \bea\hp{-9mm} L =
\frac{2}{\theta} \,\, (\dot{x}_2\, x_1 - \dot{x}_1 \, x_2 ).
\label{5.6} \eea

\noi The corresponding canonical momentum is $\,p_1
=\frac{\partial L}{\partial \dot{x}_1} = - \frac{2}{\theta}\,
x_2\,$ and \bea \hp{-7mm} [\hp{.3mm}\hat{x}_1 ,
\hat{p}_1\hp{.3mm}]  = \Big[\hat{x}_1 , - \frac{2}{\theta
}\hp{.3mm}\hat{x}_2\Big] = i\hp{.3mm} \hbar_{eff}\,. \label{5.7}
\eea

\noi In this limit one has a modified noncommutative configuration
space with $\, [\hat{x}_1 , \hat{x}_2] = - i \hbar_{eff} \,
\frac{\theta}{2} .$ The spacing between energy levels diverges
like $\,- \frac{\hbar\hp{.5mm} e^2\hp{.2mm} {\mathcal
B}^2\hp{.2mm} \theta}{4 m}\,$ and the system practically lives in
the lowest level $\,- \frac{\hbar\hp{.5mm} e^2\hp{.2mm} {\mathcal
B}^2\hp{.2mm} \theta}{8\hp{.3mm} m} ,$ which differs from the
standard Landau lowest level $\,\frac{\hbar\hp{.4mm} e
\hp{.3mm}{\mathcal B}}{2\hp{.3mm} m}\,.$

Summarizing, the main results of this paper are:

$(i)$ compact presentation of classical and quantum mechanics
related to the phase space  NC (\ref{1.4});

$(ii)$ explicit formulae (\ref{2.23}) for coefficients of
quadratic Lagrangian in noncommutative regime  with respect to the
coefficients of commutative one;

$(iii)$ general expressions  of the Schr\"odinger equation
(\ref{3.1}) and the Feynman path integral (\ref{3.4}) for
noncommutative quadratic Hamiltonian (\ref{2.14}) and Lagrangian
(\ref{2.19});

$(iv)$ Feynman path integral (\ref{4.14}) and energy spectrum
(\ref{5.5}) for a charged particle in the noncommutative plane
with perpendicular magnetic field; and

$(v)$ introduction of the effective noncommutative Planck constant
(\ref{5.2}), which could have significant cosmological
implications.

 At the end, it is worth mentioning that there are some other
commutation relations similar to (\ref{1.4}), which by linear
transformations of canonical variables can be converted to the
usual Heisenberg algebra (\ref{1.5}). This will be presented
elsewhere.

\bigskip

\section*{Acknowledgments}

The work on this article was partially supported by the Ministry of
Science and Environmental Protection, Serbia, under contract No
144032D.

\end{document}